%% file: acl_latex.tex
\pdfoutput=1

\documentclass[11pt]{article}

\usepackage[final]{acl}

\usepackage{times}
\usepackage{latexsym}
\usepackage{mdframed}

\usepackage[T1]{fontenc}

\usepackage[utf8]{inputenc}
\usepackage{subcaption}
\usepackage{microtype}

\usepackage{inconsolata}

\usepackage{graphicx}
\usepackage{color}
\usepackage{graphicx}
\usepackage{pifont}
\usepackage{multirow}
\usepackage{xspace}
\usepackage{caption}
\usepackage{amsmath}
\usepackage{enumitem}
\usepackage{wrapfig,lipsum}
\usepackage{diagbox}
\usepackage{makecell}
\usepackage{float}

\usepackage{bm}
\usepackage{framed}

\definecolor{shadecolor}{gray}{0.9}

\usepackage{tcolorbox} 
\tcbuselibrary{theorems} 
\usepackage{xcolor} 
\definecolor{mytheoremfr}{RGB}{200,200,200} 
\definecolor{mytheorembg}{RGB}{240,240,240} 

\usepackage{booktabs}
\usepackage{siunitx}
\usepackage{dashrule}
\usepackage{colortbl}

\usepackage{amsmath}
\usepackage{amssymb}
\usepackage{mathtools}
\usepackage{amsthm}

\title{EconAI: Dynamic Persona Evolution and Memory-Aware Agents
in Evolving Economic Environments}

\author{
  Annie Liu,
  Zane Cao,
  Lang Chen,
  Zongxin Xu,
  Zigan Wang \\
  Tsinghua University
}

\begin{document}
\maketitle
\begin{abstract}
The integration of large language models (LLMs) in economic simulations has significantly enhanced agent-based modeling, yet existing frameworks struggle to capture the interplay between short-term optimization and long-term strategic planning. Conventional approaches rely on static data-driven predictions, failing to incorporate adaptive behaviors influenced by economic sentiment, market volatility, and individual goals. To address these limitations, we introduce a novel EconAI framework, incorporating economic sentiment indexing (ESI), memory weighting, and dynamic decision-making mechanisms. By quantifying economic belief, adjusting historical data influence, and linking work-consumption behaviors, EconAI achieves a more human-like decision process, where agents adapt their actions based on both market signals and long-term objectives. It is the first LLM-powered simulation system that can simulate the macro/microeconomic environment and interactions in a unified framework. Empirical evaluations show that EconAI improves stability in economic responses, better replicates real-world employment-consumption cycles, and enhances overall decision robustness. This advancement marks a crucial step towards more realistic, adaptive economic agent simulations.
\end{abstract}

\input{1.intro}

\input{2.related}

\input{4.method}

\input{5.exp}

\input{Conclusion_and_Future_Work}

\bibliography{sample-base}

\end{document}

%% file: 1.intro.tex
\section{Introduction}
\label{sec::intro}

The advent of artificial intelligence (AI) \cite{huang2026radar,yin2026numcoke,cao2024diffusione} has revolutionized economic research, enabling unprecedented insights into individual behaviors, consumer preferences, and market dynamics through advanced data processing and pattern recognition~\cite{jorgenson2001information,schorfheide2015real,christiano2005nominal}. Among AI-driven methodologies, Agent-based modeling (ABM) has emerged as a powerful framework for simulating economic systems from the bottom up, capturing interactions among diverse agents without relying on predetermined equilibria~\cite{farmer2009economy}. While early ABM approaches relied on simplistic rule-based models~\cite{tesfatsion2006handbook,brock1998heterogeneous}, recent advancements have incorporated learning-based techniques to better reflect complex economic dynamics~\cite{trott2021building,zheng2022ai,mi2023taxai}. However, tailoring decision-making processes to individual agents remains a significant challenge. Customized rule sets require extensive expert knowledge and calibration~\cite{windrum2007empirical}, while neural network-based approaches often face prohibitive computational costs and training complexities~\cite{mi2023taxai}. These limitations hinder the practical application of ABM and restrict its ability to capture the full diversity of economic dynamics.

The integration of large language models (LLMs) into economic simulations has significantly advanced agent-based modeling by enhancing reasoning and planning capabilities~\cite{zhao2023survey}. However, existing frameworks often fall short in capturing the nuanced interplay between short-term optimization and long-term strategic planning, which is critical for realistic economic decision-making. A key limitation lies in their inability to effectively integrate two essential components: event memory and economic beliefs. In real-world economic environments, decision-makers rely on historical experiences (event memory) and adapt their actions based on evolving market conditions and their confidence in the economic outlook. Current LLM-driven approaches, however, often neglect these aspects, leading to an accumulation of simulation errors that can render the model increasingly unrealistic over time~\cite{DBLP:conf/acl/YueZJ24}. This gap highlights a critical limitation of current methods: their inability to simulate the dynamic interplay between memory-driven learning and confidence-driven adaptation, which is essential for accurate long-term economic modeling.

In light of these limitations, this paper addresses the following question:
\textit{How can we develop a new system that can more accurately simulate macro/microeconomic dynamics in a unified framework?}

To this end, we introduce EconAI, a novel framework designed to enhance the long-term decision-making capabilities of LLMs in economic simulations. Our framework categorizes agents into two types: households for microeconomic analysis and firms for macroeconomic perspectives. It also integrates the roles of government and financial institutions, recognizing their substantial impact on macroeconomic conditions. EconAI incorporates a knowledge memory module for each agent type, constructed from historical interactions and knowledge learned through LLMs. This memory is organized into long-term and short-term banks, capturing enduring economic patterns and immediate market conditions, respectively. The long-term memory bank stores vector representations of high-level event summaries, while the short-term memory bank retains contextual information for ongoing activities. A dedicated event memory perception module helps maintain coherence across sessions by effectively managing these memory banks, ensuring decisions are grounded in both historical context and current conditions. 

Furthermore, EconAI introduces a persona extraction module that models the evolving preferences and economic beliefs of agents. These dynamic personas are continuously updated and stored in a long-term persona bank, allowing for personalized and context-sensitive decision-making. By integrating long-term memory, dynamic persona modeling, and real-time updates, EconAI enables agents to respond to immediate economic fluctuations while considering past experiences and adjusting their strategies according to evolving economic beliefs. Our experiments demonstrate that EconAI significantly improves the accuracy of traditional economic indicators such as inflation and unemployment rates, outperforming conventional rule-based or machine-learning agents. This highlights the framework’s ability to balance short-term responsiveness with long-term strategic thinking, providing a more robust platform for economic simulations.

In summary, our contributions are three-fold:

\begin{itemize}[leftmargin=*]
\item We introduce EconAI, the first LLM-powered simulation system that integrates both macroeconomic dynamics and micro-level agent interactions within a unified framework. This allows agents to adapt to both market environments and individual economic decisions effectively.

\item EconAI incorporates multiple adaptive mechanisms, including the Economic Sentiment Index (ESI), memory weighting, and dynamic decision-making models. These features enable it to continuously adjust to market conditions and simulate real-world economic fluctuations, surpassing traditional static prediction models.

\item Through extensive macroeconomic and microeconomic simulations, we demonstrate that EconAI significantly outperforms existing models, and EconAI simulates economic indicators such as inflation and unemployment with greater precision and verifies different economic laws.

\end{itemize}

%% file: 2.related.tex
\section{Related Work}
\label{sec::related}

\subsection{Simulation in Macroeconomics}\label{sec::relatedwork1}
Macroeconomic modelling has long been caught between analytical tractability and behavioural realism. Closed-form frameworks such as DSGE~\cite{christiano2005nominal} and reduced-form empirical models~\cite{hendry1982formulation,phelps1967phillips,kydland1982time} impose strong rationality and equilibrium assumptions, which limits their capacity to represent heterogeneous agents, frictions, and out-of-equilibrium dynamics. Agent-Based Modelling (ABM) sidesteps this constraint by letting macro-level regularities emerge from the interaction of many micro-level actors, and has therefore become the preferred vehicle for policy experiments whose outcomes depend on distributional shifts rather than representative-agent averages.

Within the ABM tradition, two design philosophies dominate. Hand-crafted rule-based agents~\cite{tesfatsion2006handbook,brock1998heterogeneous} encode economists' prior knowledge directly but scale poorly as the behavioural space grows, requiring expert re-calibration for every new setting. Learning-based agents trained with deep or reinforcement learning~\cite{trott2021building,zheng2022ai,mi2023taxai} relax this rigidity but transfer the burden to data and compute, neither of which is easy to come by at the macro scale. EconAI departs from this dichotomy by grounding agent cognition in language-level reasoning, enabling a single backbone to absorb macroeconomic context and act heterogeneously across households and firms without hand-written rules or task-specific training.

\subsection{LLM-empowered Agents}\label{sec::relatedwork2}
A rapidly growing literature turns large language models into autonomous simulation agents~\cite{wang2023survey,xi2023rise}, exploiting their ability to follow natural-language instructions, retain conversational context, and recover gracefully from under-specified inputs~\cite{autogpt,babyagi}. These capabilities have been put to use in generative-agent societies~\cite{park2023generative,park2022social}, multi-agent social-science testbeds~\cite{kovavc2023socialai,gao2023s,jinxin2023cgmi,gilbert2005simulation}, and tool-using scientific assistants~\cite{boiko2023emergent,bran2023chemcrow}. The economic applications that are closest to our setting range from probing the rationality of individual LLM decisions~\cite{horton2023large,chen2023emergence}, through game-theoretic interaction between small numbers of agents~\cite{guo2023gpt,akata2023playing}, to market-scale simulation with competing firms or traders~\cite{zhao2023competeai,chen2023put}.

The work most directly comparable to ours is EconAgent~\cite{li2024econagent}, which uses an LLM to drive single-period household choices within a macroeconomic loop. Our focus is complementary: rather than treating each month as an independent rational choice, EconAI models the longitudinal trajectory of a household, tightly coupling memory retention, belief revision, and sentiment-modulated preferences inside one decision module, so that multi-year shocks and regime shifts are absorbed rather than reset at every step.

%% file: 4.method.tex
\begin{figure*}[t]
\centering
    \includegraphics[width=0.99\linewidth]{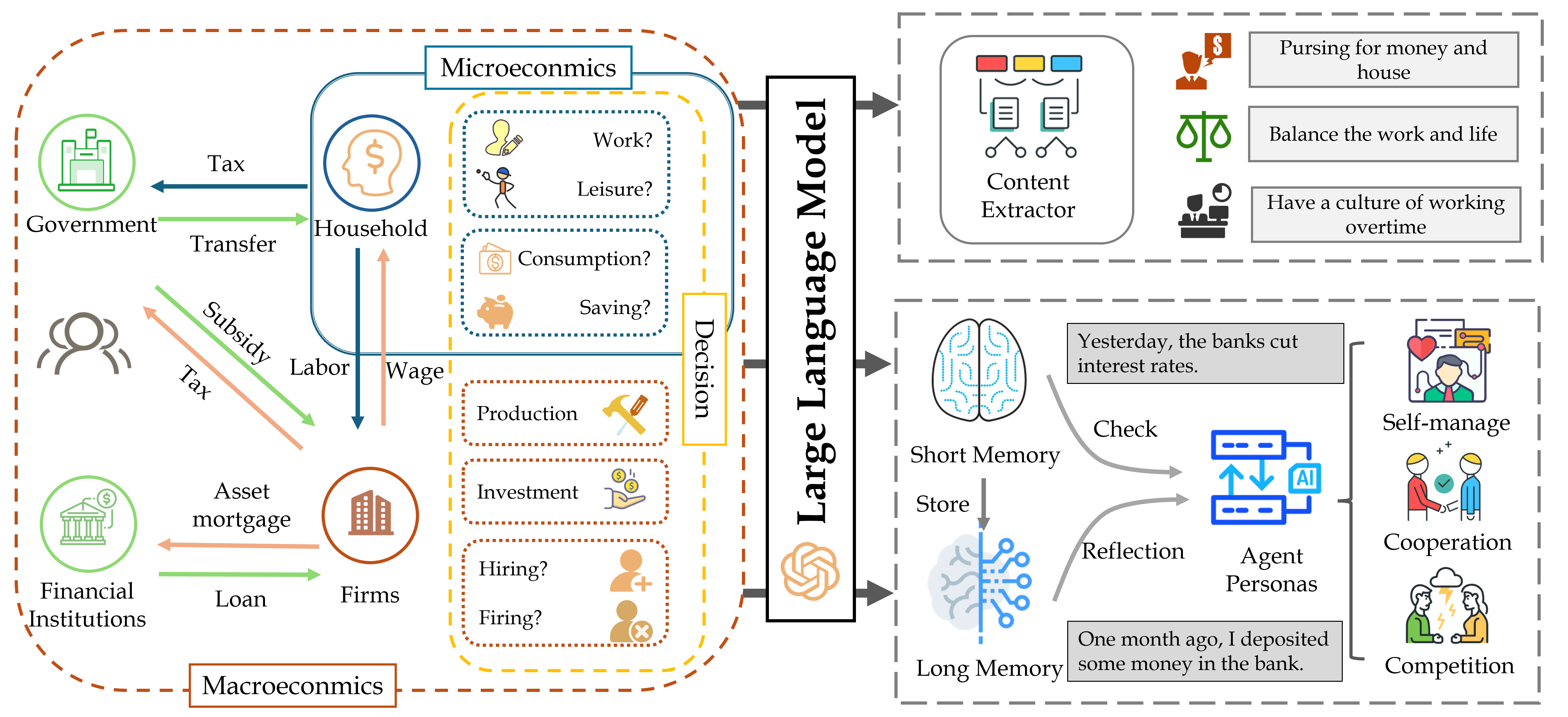}
\caption{The illustration of the simulation for the microeconomic and macroeconomic environments (left) and our EconAI (right). On the left, the simulation analyzes household microeconomic 
decisions on work and leisure, alongside firms' macroeconomic choices on production, investment, and 
employment, shaped by simplified interactions with government and financial institutions. On the right, the event 
module retains historical data in long-term memory and contextual information in short-term memory. The 
persona module dynamically extracts and updates user and agent personas from ongoing interactions, storing 
them in a dedicated personal bank. The response module integrates this information to generate contextually 
informed and coherent responses, enabling the system to support economic decision-making and simulate 
human-like behaviors such as self-management, cooperation, and competition.}
\label{fig:framework}
\end{figure*}
\section{EconAI Framework}\label{sec::method}
In this section, we introduce EconAI, which simulates the economic activity plan in the decision-making process. This plan comprises a series of abstract actions to be carried out across various scenarios with a long-term-based framework. According to economists \cite{falk2018global}, long-term decision-making rather than short-term counterparts is the base for a robust economic system. To this end, as illustrated in Figure \ref{fig:framework}, we develop a new approach to simulate the long-term decision-making process including event perception, dynamic personas extraction, and response generation for decision-making, thus conducting the economic simulation.

\subsection{Event Perception}
\label{sec:event}
The overall pipeline is illustrated in Figure~\ref{fig:framework}. We now turn to the agent-side architecture that drives each household and firm inside EconAI.

For an agent, such as a household, it has specified metadata such as the profession, specialty, skills, credentials, and experiences of the agent. The agent observes information from the environment, makes decisions, and conducts the appropriate action.  In real-world economic activities, humans often make decisions based on heuristic rules and established customs derived from experience, such as the belief that lower bank interest rates encourage investment. Afterward, people reinforce and apply these rules of thumb based on past successes or revise their heuristic rules in response to failures in specific contexts. Much like humans, the agent's brain serves as a central nucleus driven by an LLM. The brain module enables the agent to exhibit sophisticated cognitive abilities critical for professional-grade performance, including memory, planning, and reasoning.

\paragraph{Memory Storage.} The event memory module is designed to perceive historical events to generate coherent responses across interval time. 
As shown in Figure~\ref{fig:framework}, 
this event memory module is segmented into two major sub-modules that focus separately on long-term and short-term memory.
The memory module is designed to capture and encode events from previous interactions in economic activities. This process involves logging both the timestamps $t$ and concise summaries $o$, which are transformed into representations and stored in a low-cost memory bank $M_L=\{\phi(t_j, o_j) \mid j \in \{1,2, \dots, l\}\}$. Here, $\phi(\cdot)$ denotes the text encoder (such as MiniLM~\cite{DBLP:conf/nips/WangW0B0020}), while $l$ represents the size of the memory bank. The encoded information is then made easily accessible via an embedding-based retrieval mechanism, facilitating efficient access to stored memories.

\paragraph{Event Summary.}
Unlike previous agent methods~\cite{DBLP:conf/uist/ParkOCMLB23} that depend solely on the zero-shot capabilities of LLMs for event extraction and summarization, we incorporate instruction tuning~\cite{DBLP:conf/iclr/WeiBZGYLDDL22} into our event summary module to enhance the quality of the summaries for economic activities. We remember the event in a structured format consisting of: (1) a task background introduction, (2) relevant economic activities to be comprehended, and (3) specific summarization instructions. These elements serve as input prompts and generate the output. The event summary module is fine-tuned using this data, effectively improving its summarization capabilities.

\subsection{Economic Preference and Confidence}
Memory information is objective, but human decision-making is often affected by emotions and confidence, such as being overly pessimistic during an economic recession and overly optimistic during an overheated economy. Equal weighting of short-term and long-term information cannot reflect people's different sensitivities to economic changes in different time frames. Then, we introduce the Economic Sentiment Index (ESI) to quantify the economic belief of agents and store this ESI value in the memory module:
\begin{equation}
  \text{ESI}_t = \lambda \text{ESI}_{t-1} + (1 - \lambda) \text{ESI}_{t,\text{LLM}}   
\end{equation}
Where $\lambda$ is the memory decay factor (0.8~0.95).
$\text{ESI}_{t,\text{LLM}}$ is the latest economic sentiment index generated by LLM.

In this way, the economic sentiment of each quarter can be quantified and stored for subsequent decision-making calculations. In this way, we can make the results of LLM reflection quantifiable, reducing computational overhead. Incorporating the dynamic changes in economic belief makes the decision-making of the intelligent agent not only data-dependent but also emotionally influenced. The impact of historical economic data will decay, avoiding excessive reactions by the intelligent agent due to short-term economic shocks.

Then, we use confidence to adjust the final decision.
If confidence is low (e.g., below 0.5). Adjust the final decision in conjunction with the economic sentiment index:
\begin{equation}
    p_{w_i} = \sigma(p_{w_i} - \theta \text{ESI}_t) , p_{c_i} = \sigma(p_{c_i} + \beta \text{ESI}_t)
\end{equation}
If economic belief is low ($\text{ESI}_t < 0$), the intelligent agent reduces consumption and increases the willingness to work. If economic belief is high ($\text{ESI}_t > 0$), the intelligent agent increases consumption and reduces work.

\subsection{Economic Environment Design}
This section outlines the design and decision-making processes of intelligent economic agents, government-imposed taxation policies, production-consumption dynamics, and financial market mechanisms shaping macroeconomic behavior. 
These components interact within a unified framework that captures real-world economic principles.
 
\noindent\textbf{Agent Decision-Making Framework.}
The core decision-making process for each agent involves two key economic choices: work and consumption. Initially, these decisions are modeled as:
\begin{equation}
    p_{w_t}, p_{c_t} \sim \text{LLM}(z_i, P, s_i, u, r,\text{ESI}_t)
\end{equation}
where \( p_{w_t} \) is the labor market participation probability, and \( p_{c_t} \) is the consumption-to-income ratio. \( p_{w_i} \) denotes the agent's willingness to work, determining whether they choose employment, while \( p_{c_i} \) represents the agent's consumption propensity, indicating the proportion of income allocated to spending. The variable \( z_i \) corresponds to the agent's individual income in the current period, while \( P \) reflects the prevailing market price of essential goods. \( s_i \) accounts for the agent’s current savings, which influence financial security and spending capacity. The unemployment rate \( u \) provides insight into labor market conditions, affecting job availability and economic confidence. Lastly, \( r \) represents the bank interest rate, which impacts savings growth and borrowing costs. These inputs collectively allow the LLM to generate dynamic and context-aware decisions that align with real-world economic behaviors.

\noindent\textbf{Production and Consumption Dynamics.}
Labor supply contributes to goods production following a Cobb-Douglas function:
\begin{equation}
    G \gets G + S = G + A K_t^{\beta_1} L_t^{\beta_2} e^{-\lambda K_t},
\end{equation}
where \( A \) represents the overall productivity factor, \( K_t \) is the capital stock, and \( L_t = \sum_{j=1}^{N} l_j \times 168 \) denotes the total labor hours supplied. The diminishing returns parameter \( \lambda \) ensures that excessive capital accumulation does not indefinitely increase output.
Capital evolves dynamically as:
\begin{equation}
    K_t = (1 - \delta) K_{t-1} + I - \omega K_t,
\end{equation}
where \( \delta \) represents depreciation, \( I \) is new investment, and \( \omega K_t \) accounts for inefficiencies.
Aggregate goods demand is determined by$ D = \sum_{j=1}^{N} d_j = \sum_{j=1}^{N} \frac{q_j^c s_j}{P}$,
where \( q_j^c \) represents the consumption propensity of agent \( j \), \( s_j \) is the savings level, and \( P \) is the market price of goods.Then, the market imbalances is defined as:
$\hat{\varphi} = \frac{D - G}{\max(D, G)}$. It
drives price and wage adjustments:
\begin{equation}
    w_i \gets w_i (1 + \varphi_i \cdot \kappa), \quad P \gets P (1 + \varphi_P \cdot \kappa),
\end{equation}
where \( \varphi_i \sim \text{sign}(\hat{\varphi}) U(0, \beta_w |\hat{\varphi}|) \) influences wages, and so does $\varphi_p$.

\noindent\textbf{Government Taxation and Redistribution.} To enhance economic stability and equity, the government implements a progressive taxation system, 
where tax rates rise with income. Post-tax income incorporates redistributed tax revenues, promoting 
partial wealth reallocation and reducing extreme income inequality. In financial markets, agents' savings 
grow at an interest rate influenced by unemployment and inflation dynamics. This rate follows a modified 
Taylor rule, adjusting for deviations from target inflation and natural unemployment. Detailed formulations 
and mechanisms are provided in the Appendix.

\subsection{Response Generation}
\label{sec:response}
When the agent engages in economic activity, it will recall the memory module, which combines event extractor: retrieved relevant memories $m$, the short-term memory context $M_S$, and the personas $P_a$ representing the agent. These inputs are passed to a response generator, which produces an appropriate response $r$. To improve the agent's capability to generate coherent and contextually relevant responses for economic activities,  for each example spanning a series of times (such as several months in the simulation), we dynamically simulate the economic activity's progression. As new economic events  arise, the system applies event summarization, persona extraction, and  topic-aware memory retrieval to collect relevant context and retrieve 
associated memories and personas for the user and agent.
This rich data is then incorporated into a response generation prompt (details in Appendix). On top of this, the agent can summarize the satisfied score for the current state and then make the decisions.

%% file: 5.exp.tex
\section{Experiments}\label{sec::exp}
In this section, we conduct experiments to study the ability of EconAI, aiming to answer the following research questions (RQ). RQ1: How does EconAI perform in simulations compared to conventional models? RQ2: What role do the key components of EconAI play in influencing the simulation outcomes? RQ3: Is the decision-making process within EconAI interpretable, and do simulations effectively capture the impact of external interventions?

\subsection{Experimental Setup}

\paragraph{Baselines.} We benchmark EconAI against four representatives of the current design space. LEN~\cite{lengnick2013agent} and CATS~\cite{gatti2011macroeconomics} are long-established rule-based macro-ABMs whose hand-crafted work/consumption rules have been repeatedly validated in follow-up simulation studies, making them natural yard-sticks for the rule-based end of the spectrum. To approximate the heterogeneity that real populations exhibit, we additionally report a Composite variant in which each synthetic agent draws its rule from the union of the LEN and CATS rulebooks at initialisation and holds it fixed thereafter. On the learning-based side, AI-Eco, based on AI-Economist~\cite{zheng2022ai}, trains utility-maximising policies via reinforcement learning~\cite{arulkumaran2017deep}; it represents the optimisation-driven end of the spectrum. Finally, we compare against EconAgent~\cite{li2024econagent}, the closest LLM-based baseline, which drives per-period household decisions through a perception/reflection loop; including it allows us to isolate the contribution of EconAI's memory-weighting and sentiment-indexing mechanisms.

\paragraph{Definition of Economic Indicators.} The annual nominal GDP is calculated as the total of $S\times P$ over the course of a year. To compute real GDP, we set the first year of the simulation as the base year, replacing $P$ with $P_0$, where $P_0$ represents the price of goods in that reference year. The wage inflation rate follows a similar formula to price inflation, except the average price is substituted by the average wage across all agents. For households, disposable income refers to the total income left after taxes and essential expenditures. The savings rate is the proportion of disposable income that is saved rather than used for consumption. For firms, profit margin is defined as the ratio of net profit to total revenue, reflecting the firm's profitability.

\paragraph{Simulation Setup.} Every synthetic household is initialised with a short persona string (a name sampled from the LLM, an age, and an occupation) that is injected verbatim into the prompt the agent receives at each step, giving the model a stable reference point across months. Ages are drawn from the 2018 U.S. Census adult distribution restricted to the 18 to 60 working population~\cite{bureau_data_nodate}, while hourly earnings follow a Pareto distribution whose scale and shape parameters are calibrated so that the induced monthly income distribution and effective tax brackets match 2018 U.S. tax filings~\cite{zheng2022ai}. To inject wage dispersion into the textual layer of each persona, we ask the LLM to propose ten representative job titles per income decile and sample an occupation accordingly. Employment transitions are sticky: a household that worked in month $t$ keeps its job at $t{+}1$, whereas an unemployed household is presented with a fresh offer whose wage is resampled from the current cross-sectional distribution. Full distributional statistics are deferred to the supplementary material. All experiments use GPT-4o-mini through the OpenAI API, with simulation orchestration implemented in Python.

\subsection{Macro-level Analysis~(RQ1)}
\begin{figure}[t]
\centering
\subfloat[Inflation rate.\label{fig:inflation-rate}]{\includegraphics[width=0.48\linewidth]{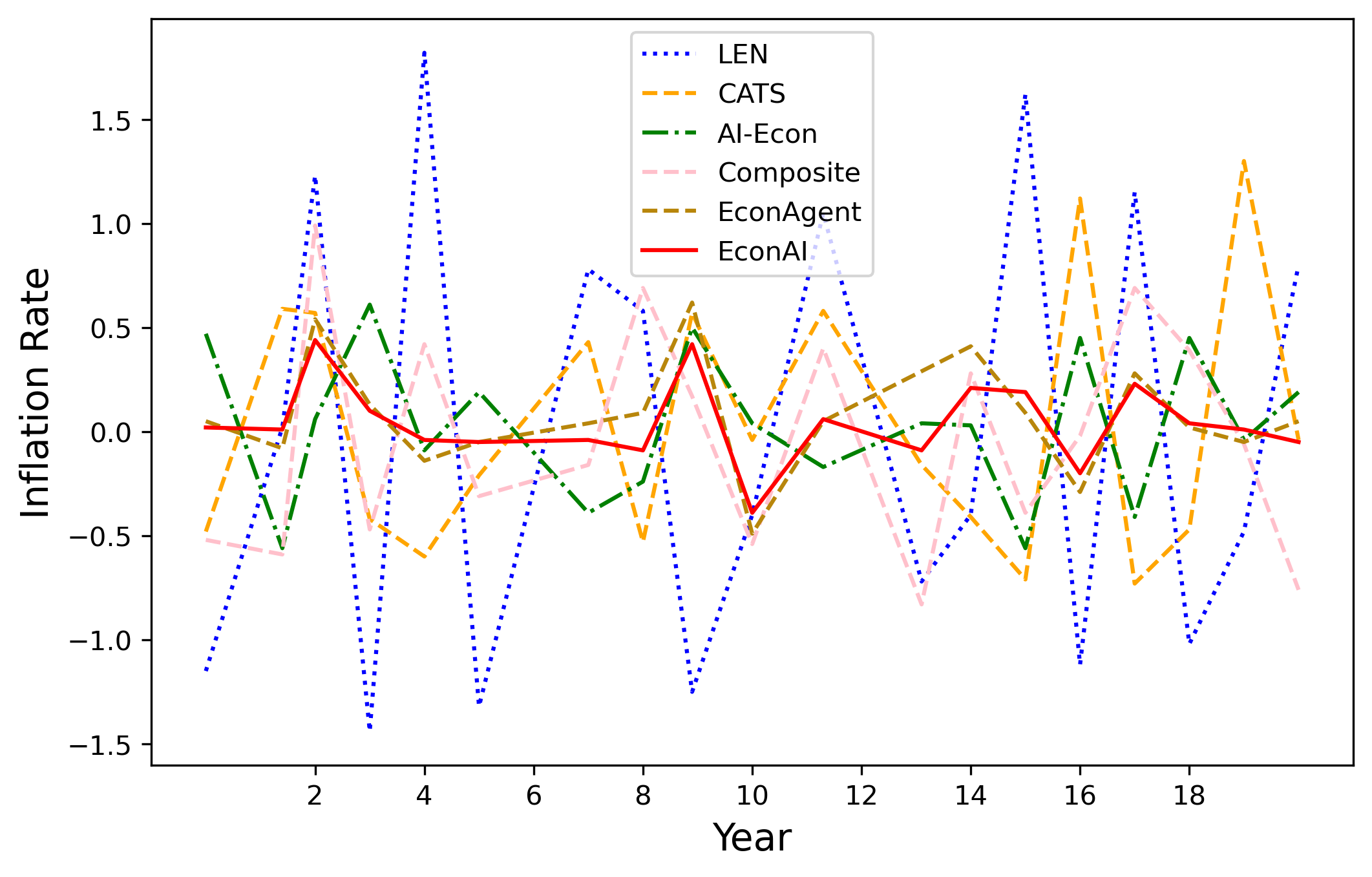}} \hfill
\subfloat[Nominal GDP Growth.\label{fig:nominal-gdp}]{\includegraphics[width=0.48\linewidth]{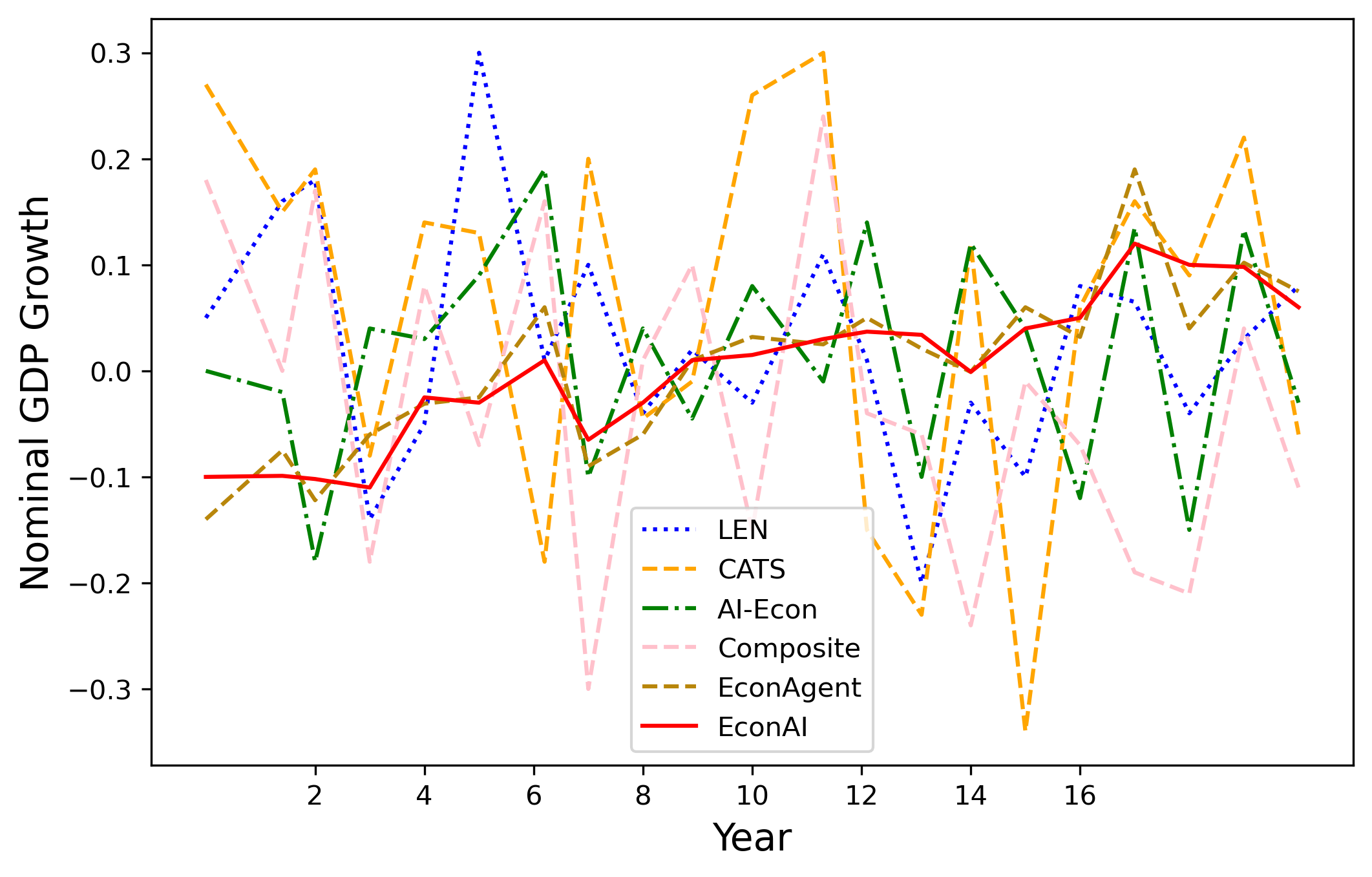}} \\
\subfloat[Normal GDP.\label{fig:unemployment-rate}]{\includegraphics[width=0.48\linewidth]{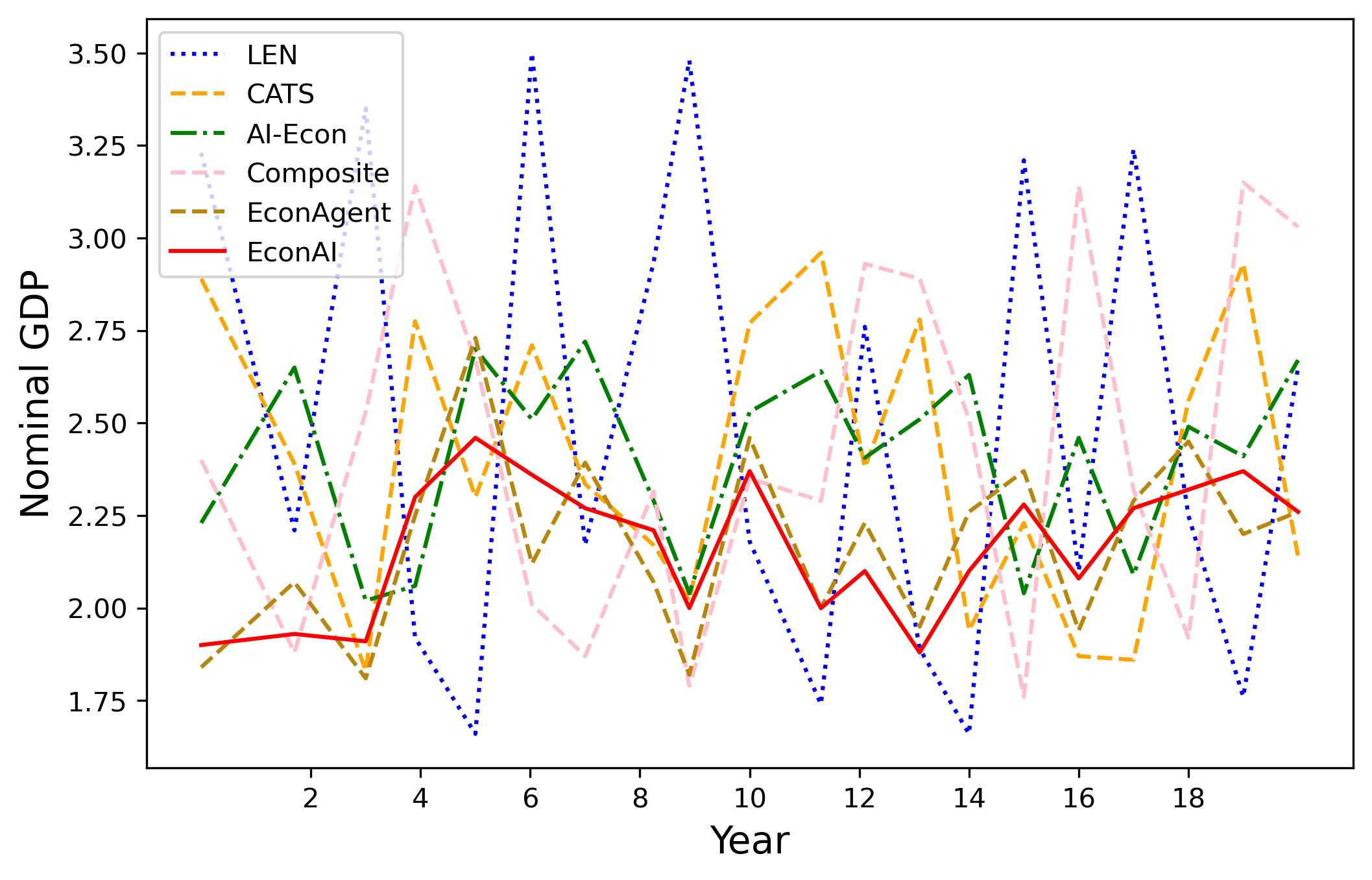}} \hfill
\subfloat[Unemployment rate.\label{fig:interest-rate}]{\includegraphics[width=0.48\linewidth]{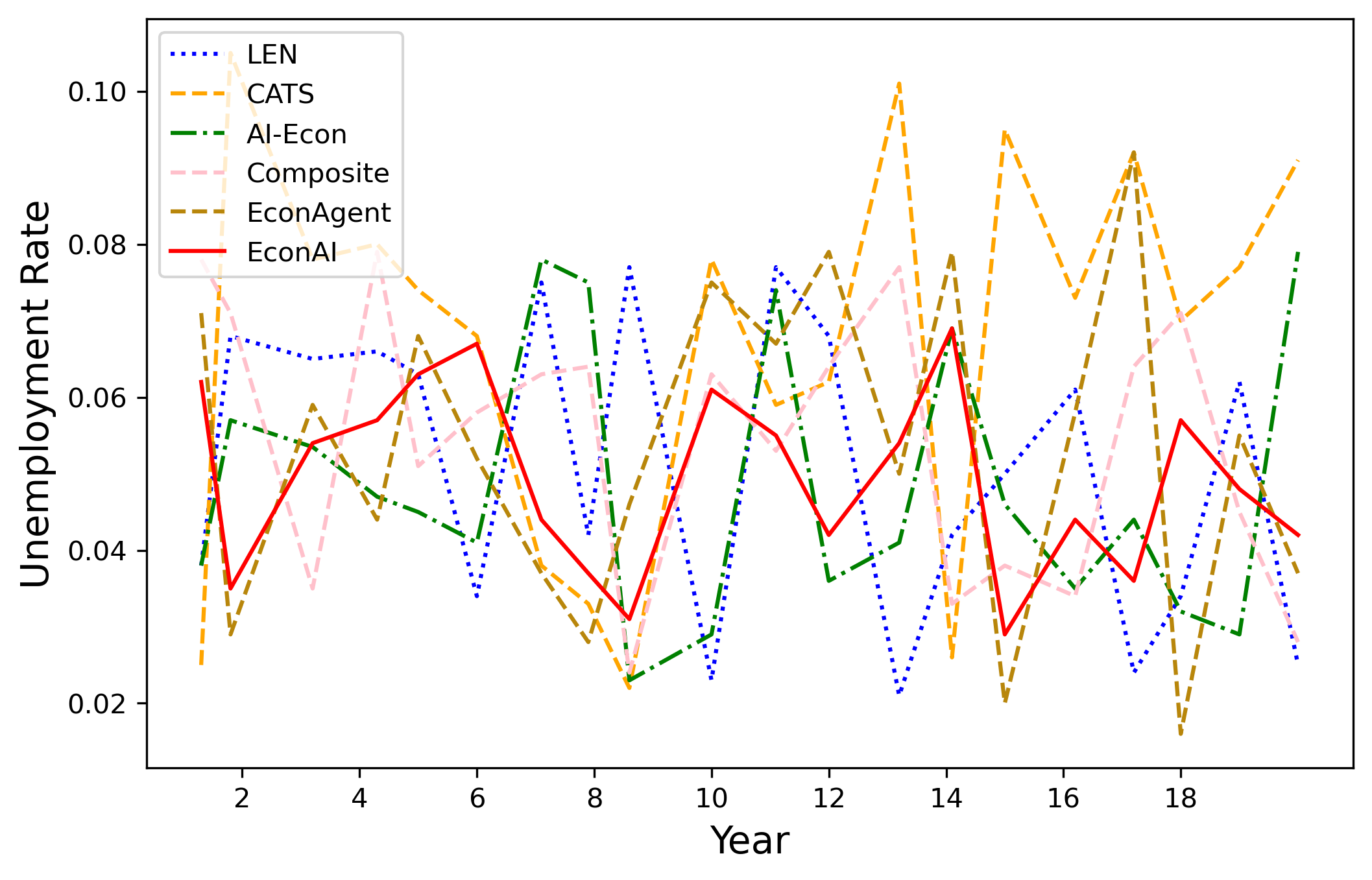}}
\caption{Annual variations of macroeconomic indicators, where the simulation based on EconAI shows more stable and numerically plausible indicators.}\label{fig::annual-indicators}
\end{figure}

\begin{figure}[t]
\center
\subfloat[Phillips curve .\label{subfig:phillips}]{\includegraphics[width=0.48\linewidth]{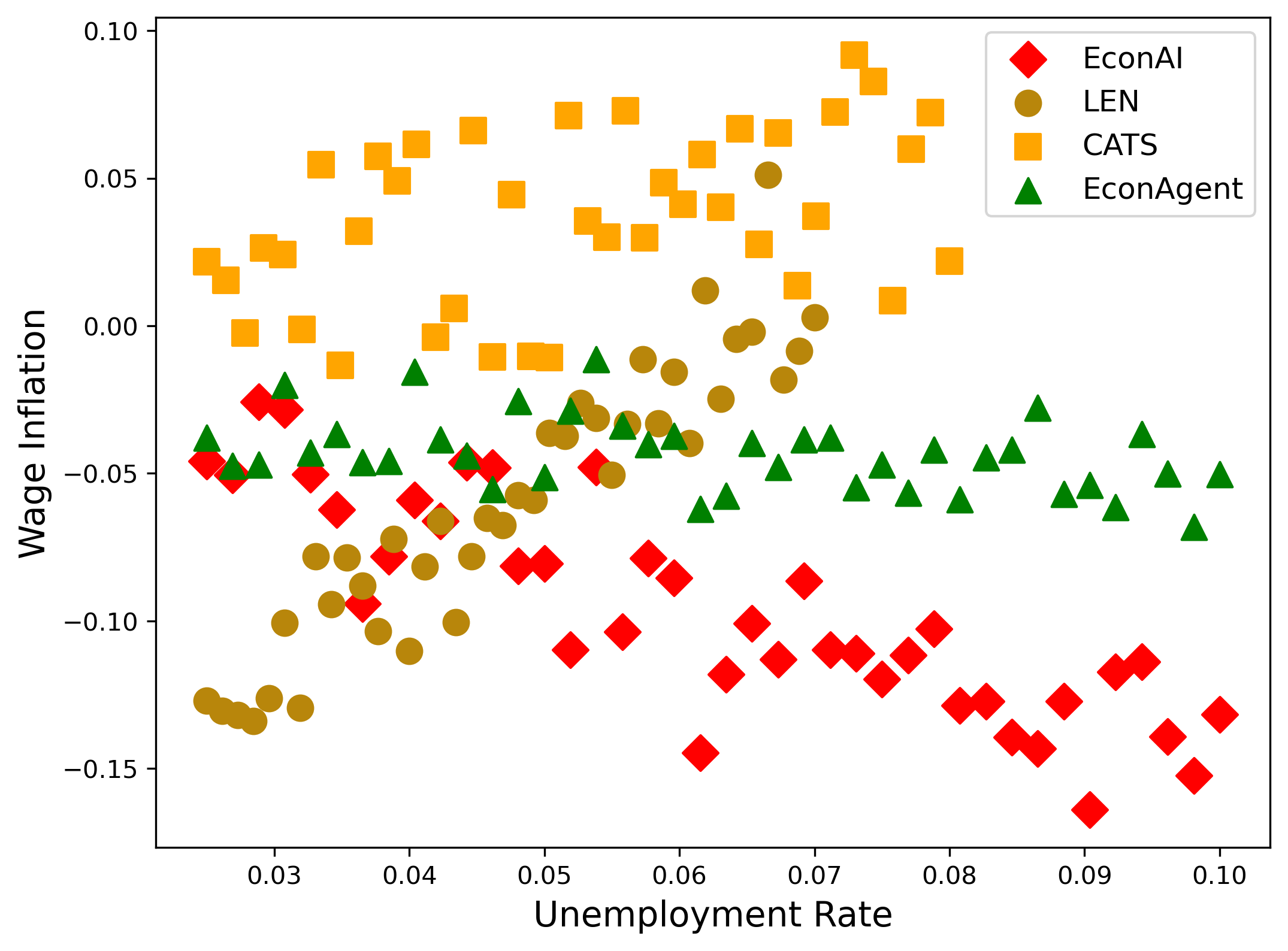}}
\subfloat[Okun curve.\label{subfig:okun}]{\includegraphics[width=0.48\linewidth]{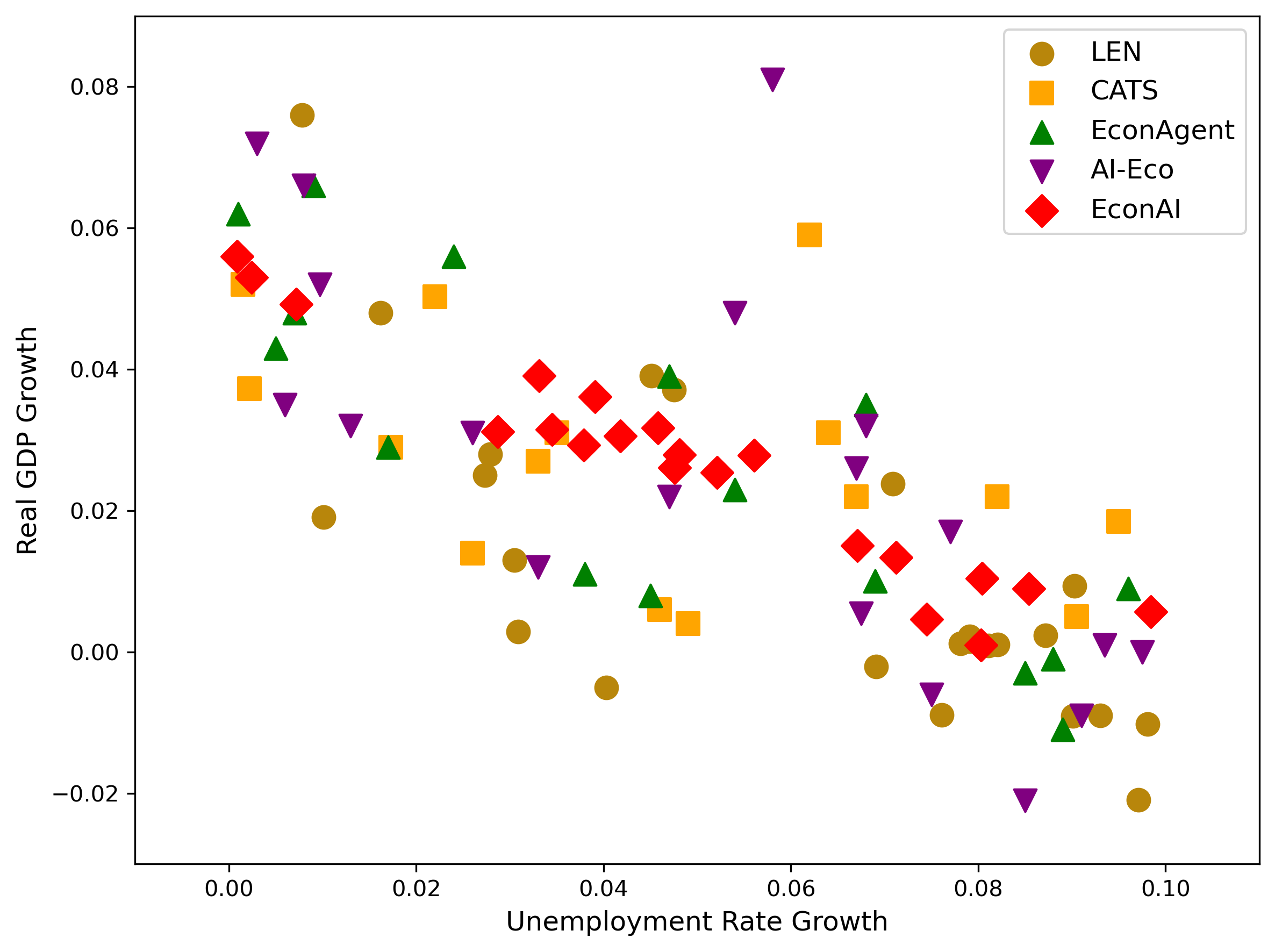}}
\caption{Economic regularity study. }\label{fig:Phillips Curve}
\end{figure}
\paragraph{Economic Indicators.} Figure~\ref{fig::annual-indicators} tracks four aggregate indicators, namely annual inflation, unemployment, nominal GDP, and nominal-GDP growth, over the 20-year horizon. AI-Eco drifts into a degenerate regime, with its unemployment rate pinned near 46\% and nominal output swinging over implausibly wide bands, so we drop it from the plots rather than let it compress the remaining curves. The rule-based and reinforcement-learning baselines share a common failure mode: every indicator shows repeated large-amplitude oscillations, a symptom of decision rules that cannot smooth short-run shocks before they propagate through the labour and consumption loop. EconAI's trajectories sit in a substantially tighter envelope without any per-variable tuning, because the memory and sentiment channels absorb transient disturbances inside the agent rather than routing them back into aggregate demand. We read this as the qualitative signature of a simulator whose micro-level decisions approximate how human households actually adjust under uncertainty.

\begin{figure}[t]
\centering
\subfloat[Competition.\label{subfig:competition}]{\includegraphics[width=0.45\linewidth]{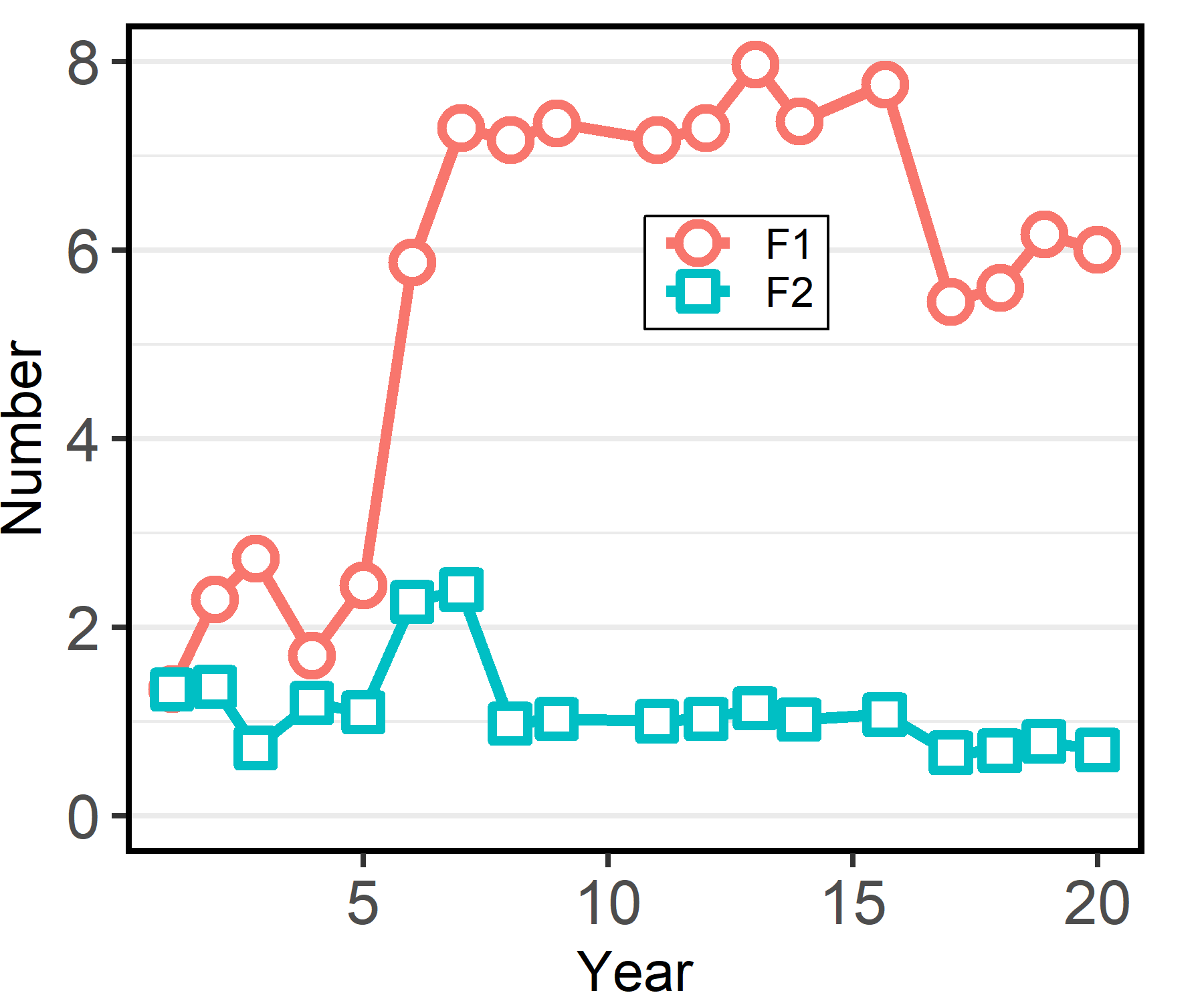}}
\subfloat[Cooperation.\label{subfig:cooperation}]{\includegraphics[width=0.45\linewidth]{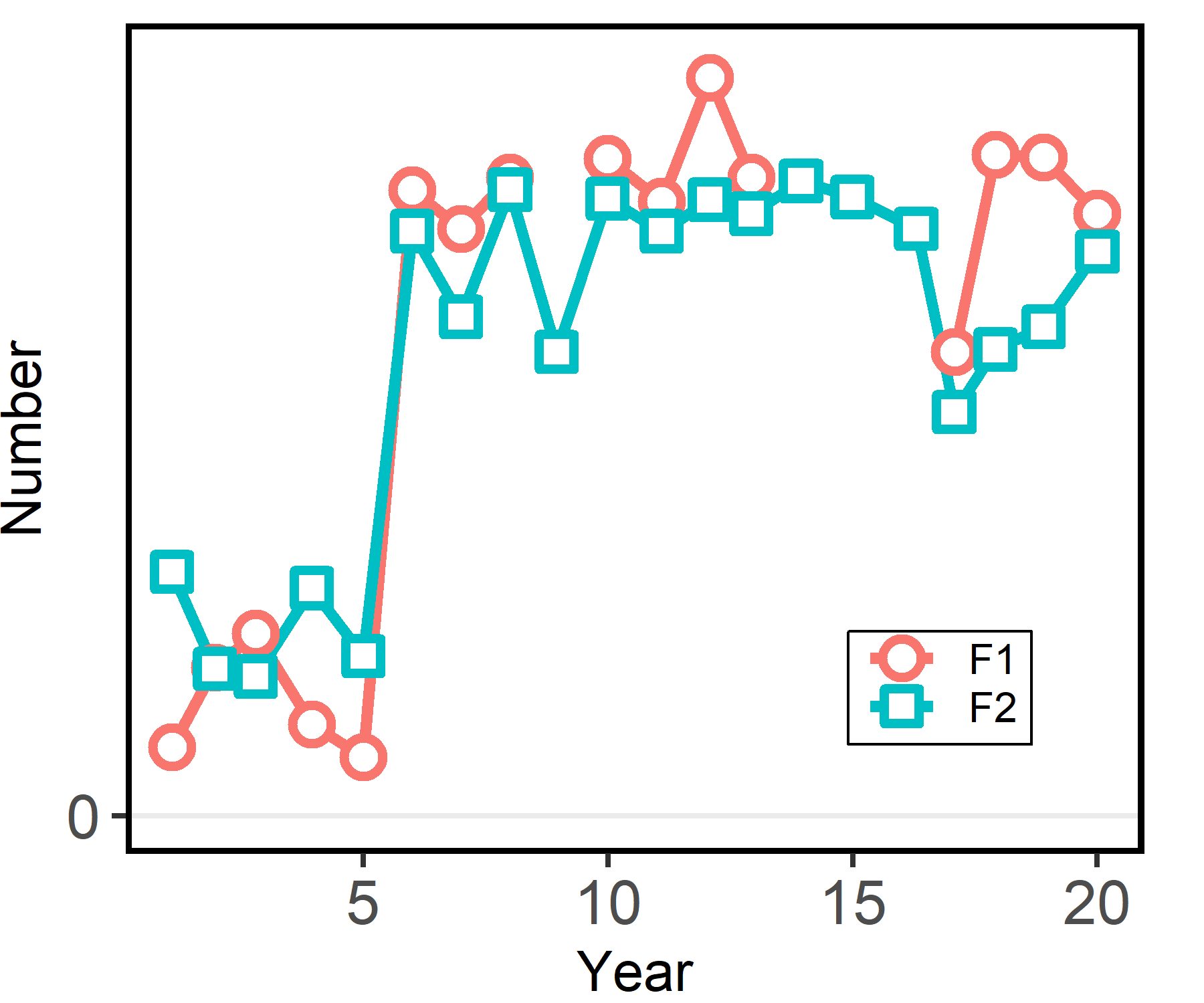}}
\caption{The experiments about the macro-level analysis. }\label{fig::The experiments about the long term and short term}
\end{figure}

\paragraph{Economic Regularity.} A stronger test of macroeconomic realism is whether simulated aggregates reproduce canonical empirical regularities. We evaluate two: the Phillips relation between unemployment and wage inflation~\cite{phelps1967phillips}, and Okun's law linking the change in unemployment to real-output growth~\cite{okun1963potential}. Figure~\ref{fig:Phillips Curve} reports both. EconAI recovers the textbook downward-sloping Phillips relation (Pearson $\rho = -0.522$, $p<0.01$) and the expected co-movement between unemployment growth and real-GDP growth under Okun's law. EconAgent and the rule-based and learning-based baselines fail at least one of the two tests, and the rule-based family in particular produces a perverse positive Phillips slope, an artefact of decision rules that do not condition consumption on the employment signal. We attribute EconAI's success to the behavioural channel inside the agent: households contract spending when their perceived unemployment risk rises, and this precautionary response propagates through aggregate demand to wages rather than being hard-coded as a slope.

\subsection{Micro-level Analysis~(RQ1)}
\label{strategy-micro-level}

\begin{figure*}[t]
\centering
\subfloat[Ablation study.\label{Ablation study}]{\includegraphics[width=0.33\linewidth]{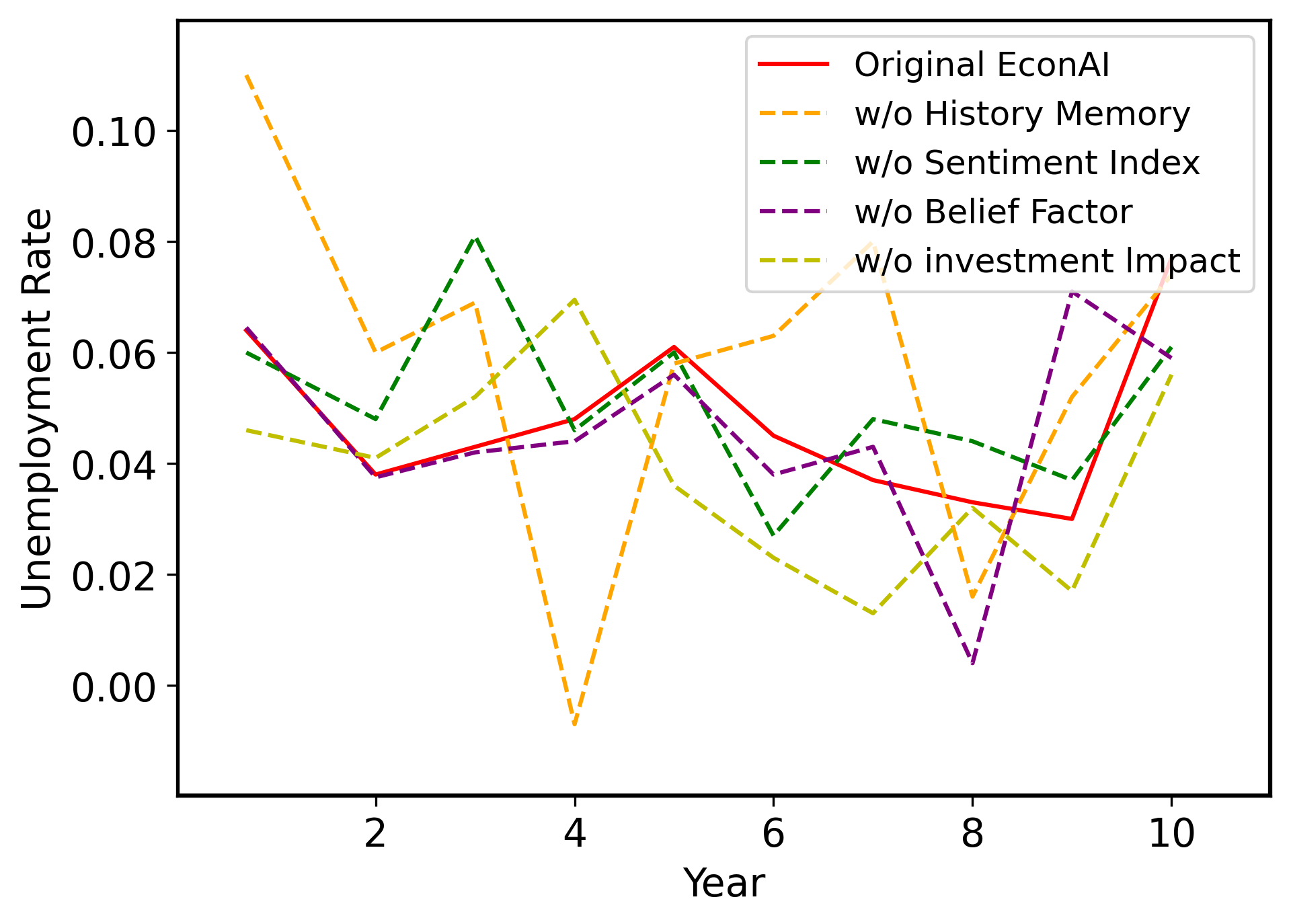}}
\subfloat[COVID-19 brings the surge of simulated unemployment rates.\label{COVID-19}]
{\includegraphics[width=0.31\linewidth]{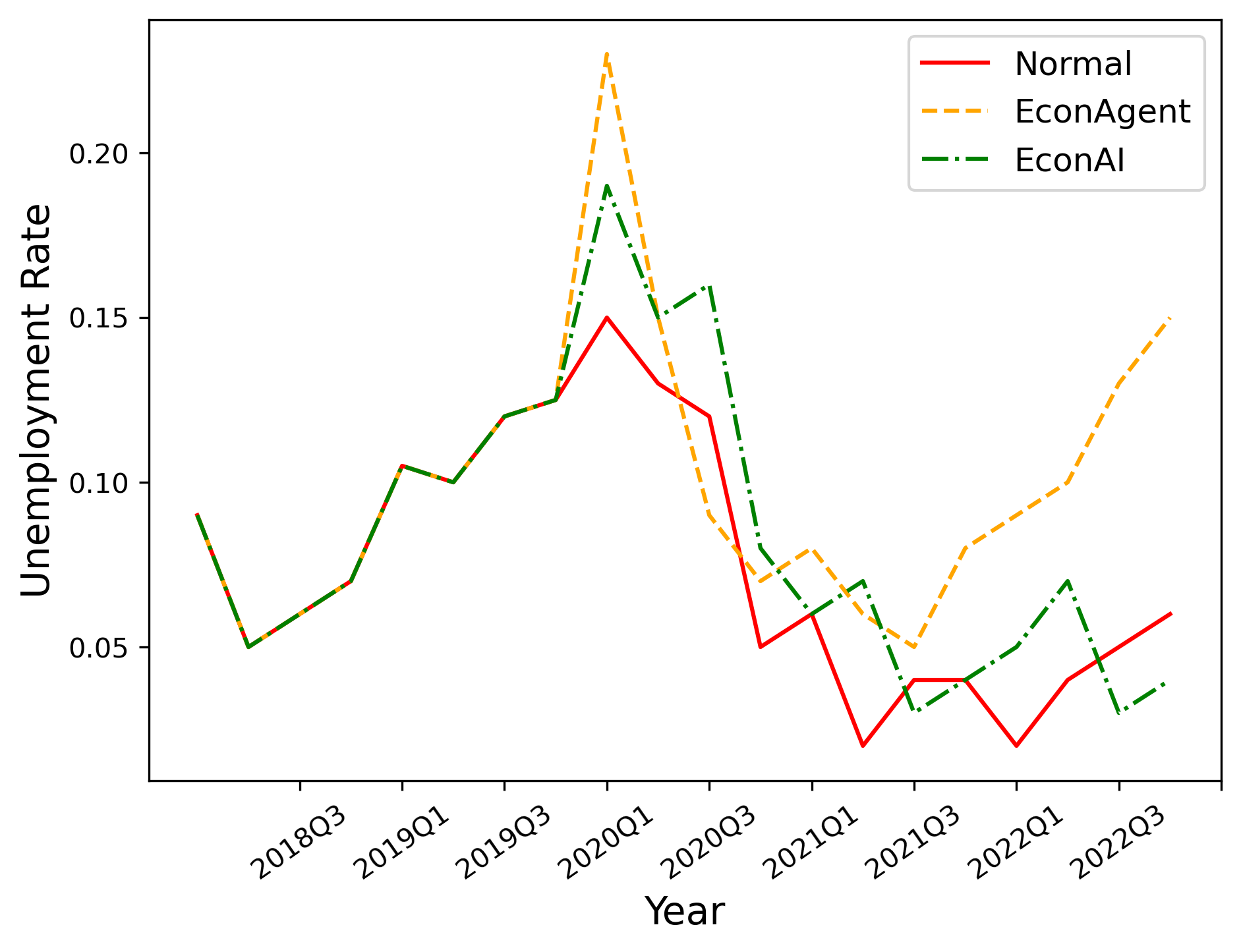}}
\subfloat[Inflation rate under different number of agents.\label{Inflation rate under different number of agents}]
{\includegraphics[width=0.33\linewidth]{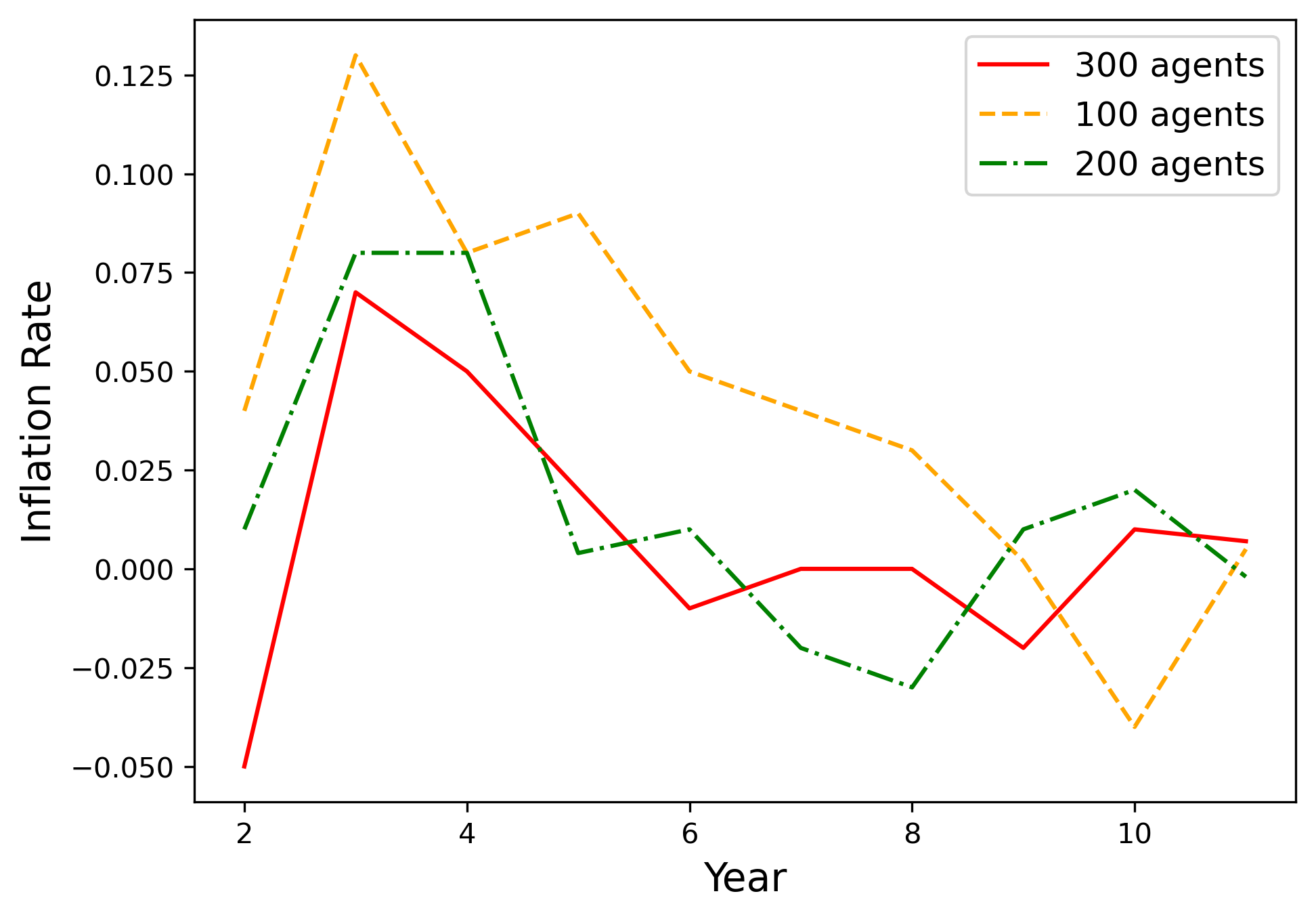}}
\caption{More experimental studies.} \label{More experimental studies}
\end{figure*}

In the economic environment modeled by EconAI, simulations of firms reveal classic market strategies and economic phenomena such as competition and cooperation. Our experiments are illustrated in Figure \ref{fig::The experiments about the long term and short term}, 
One scenario where one firm’s revenue stagnates or declines significantly suggests destructive competition. This occurs when firms engage in aggressive price wars, excessive resource allocation to outcompete rivals or strategic behaviors that undermine industry stability. In game theory, such dynamics resemble a \textit{zero-sum} or \textit{negative-sum game}, where one firm's gain comes directly at the expense of another, sometimes leading to a market exit or inefficiencies. When competition becomes excessively aggressive without value creation, the industry as a whole may suffer from reduced profitability and instability. This simulation thus provides an economically rational representation of competition, reinforcing that constructive rivalry leads to mutual growth, while destructive competition can hinder overall economic progress.

In contrast, in the second scenario, where both firms experience revenue growth, the simulation reflects principles of healthy competition. Economic theory suggests that when firms compete in ways that drive innovation, improve efficiency, and expand market demand, both can thrive. This aligns with the game-theoretic concept of positive-sum competition, where firms engage in strategic differentiation rather than direct conflict, allowing them to capture different market segments or create complementary value. By fostering innovation and efficiency rather than resorting to destructive tactics, both firms achieve sustainable long-term growth, leading to increased overall welfare in the industry.

\subsection{Ablation Study~(RQ2)} 
We conduct an ablation study by independently removing key economic factors, with the results over a 10-year span shown in Figure \ref{Ablation study}. Specifically, we analyze five versions: the original model (EconAI), removing historical memory (denoted as \textit{No History Memory}), removing the sentiment index (\textit{No Sentiment Index}), removing the belief factor (\textit{No Belief Factor}), and removing investment impact (\textit{No Investment Impact}). Notably, the absence of historical memory leads to sharp fluctuations in the unemployment rate, suggesting that past economic trends play a crucial role in stabilizing employment levels. Similarly, removing the sentiment index results in increased volatility, highlighting the influence of public sentiment on labor market dynamics. The absence of the belief factor introduces inconsistencies, reinforcing its role in guiding employment expectations. Finally, removing investment impact leads to overall lower unemployment variability, suggesting that investment plays a stabilizing role. These findings emphasize the importance of incorporating these components.

\subsection{External Intervention~(RQ3)}
To probe whether EconAI responds plausibly to an exogenous regime shift, we inject a single textual event (the March 2020 declaration of a COVID-19 national emergency in the United States) into every agent's prompt from that month onward and leave every other parameter untouched~\cite{dawid2018agent}. Figure~\ref{COVID-19} contrasts the resulting unemployment trajectory against a counterfactual run without the injection. Simulated unemployment spikes sharply in early 2020, recovers only partially, and remains elevated relative to the counterfactual through 2021, consistent with the lagged labour-market recovery documented by \cite{organization_for_economic}, even though the agents are given nothing more quantitative than a qualitative cue. Inspection of the reflection traces shows households shifting toward precautionary savings and reasoning explicitly about income diversification, which is the behavioural channel through which the prompt is converted into aggregate effects. We therefore read the experiment as evidence that the prompt interface is a sufficient conduit for macro-level shocks; numerical exactness is not the intended deliverable. 
\paragraph{Sensitivity and Robustness} We increase the number of agents to 300 and rerun the simulation. As illustrated in Figure \ref{Inflation rate under different number of agents}, the inflation rates remain consistently stable and realistic, closely resembling the results obtained with 100 agents. This trend is also observed across other economic indicators, indicating that the simulation outcomes are not significantly affected by changes in the number of agents.

%% file: Conclusion_and_Future_Work.tex
\section{Conclusion}
We presented EconAI, an LLM-driven agent-based macroeconomic simulator whose household and firm sides share a single cognitive backbone composed of event perception, a weighted long-term/short-term memory bank, and a sentiment-modulated response module. The key departure from prior LLM-agent work is the Economic Sentiment Index together with an explicit weighting between remembered history and current context, which together let an agent modulate its work--consumption decisions according to its evolving belief about the environment rather than a static rule or a single-period optimum. Over a 20-year simulation, this design is what keeps inflation, unemployment, and nominal GDP inside empirically plausible envelopes; it is also what lets EconAI recover the downward Phillips slope and Okun's co-movement on which the rule-based and learning-based baselines fail, and convert a single qualitative COVID-19 prompt into a quantitatively coherent labour-market response. The broader take-away is that binding memory, sentiment, and language-level reasoning into one loop is a viable substitute for hand-engineered rulebooks or bespoke training pipelines in macroeconomic agent design.

\section{Limitation}
The proposed EconAI framework demonstrates promising advancements by integrating event memory, economic sentiment, and dynamic persona extraction to bridge short-term optimization and long-term strategic planning. However, several limitations remain. First, its reliance on large-scale language models results in significant computational and storage overhead, potentially impeding scalability and real-time applicability in extensive simulations. Second, the complexity of integrating multiple modules introduces challenges in parameter tuning, with key parameters such as the memory decay factor and sentiment weighting requiring further validation across diverse economic scenarios. Finally, the experimental evaluations are primarily based on simulated data, which may not fully capture the nuances of all real-world economic fluctuations and extreme events. Addressing these limitations will enhance the model's robustness and broaden its practical applicability.